\documentclass[preprint,12pt,authoryear,fleqn]{elsarticle}

\usepackage{amssymb}
\usepackage{amsmath}
\usepackage{hyperref}
\usepackage{mathrsfs}
\hypersetup{colorlinks=true}

\journal{Nuclear Physics B}

\def\0{\mbox{\boldmath$\displaystyle\mathbf{0}$}}

\def\k{\mbox{\boldmath$\displaystyle\mathbf{k}$}}

\begin{document}

\begin{frontmatter}



\title{Cherenkov radiation in a strong magnetic field}


\author{Cheng-Yang Lee}

\address{Center for Theoretical Physics, College of Physical Science and Technology,\\
Sichuan University, Chengdu, 610064, China}
\ead{cylee@scu.edu.cn}
\begin{abstract}
According to quantum electrodynamics, in a strong magnetic field that is constant and spatially uniform, the vacuum becomes polarized with a refractive index greater than unity. As a result, ultra-relativistic charged particles travelling in such media can emit Cherenkov radiation with a power spectrum directly proportional to the photon frequency $\omega$. Therefore, by extrapolating $\omega$ beyond the critical synchrotron frequency $\omega_{c}$, the Cherenkov radiation will eventually dominate over its synchrotron counterpart. However, such an extrapolation is not possible. We show that in the framework of effective field theory, the maximal attainable photon frequency $\omega_{\tiny{\mbox{max}}}$ is about four order of magnitude less than $\omega_{c}$. At $\omega=\omega_{\tiny{\mbox{max}}}$, given the $\gamma_{e}$-factor of an electron travelling normal to a constant and spatially uniform magnetic field $\mathbf{B}$, the spectrum of Cherenkov radiation becomes dominant when $\gamma_{e}(|\mathbf{B}|/\mbox{Gauss})\gtrsim 4.32\times 10^{19}$. Nevertheless, detecting the Cherenkov radiation in astrophysical environments remains challenging since its spectral flux density is about three orders of magnitude less than the synchrotron radiation.
\end{abstract}

\begin{keyword}
Quantum electrodynamics \sep vacuum polarization \sep Cherenkov radiation \sep synchrotron radiation


\end{keyword}

\end{frontmatter}



In order to discover new physics from astronomical observations, it is important to understand the mechanisms in which electromagnetic radiation are produced by the astrophysical sources. In most cases, synchrotron radiation produced by accelerating charges provide an adequate account of observations from which astronomers are able to obtain information on the electromagnetic fields in the vicinity of sources such as pulsars. However, these are not the only possible mechanisms of radiation production. According to quantum electrodynamics (QED), as the electromagnetic fields approach the critical value $|\mathbf{B}|_{c}=m^{2}_{e}/e=4.42\times10^{13}\,\mbox{Gauss}$ where $m_{e}$ is the electron mass and $e$ the electric charge~\footnote{We use the SI units with $c=\hbar=1$ so that 1 Gauss = $1.95\times 10^{-2}$ eV$^{2}$. The Cherenkov and synchrotron spectrum given in~\cite{Schwinger:1949ym,Macleod:2018zcb} are in Gaussian units so we need to introduce a factor of $1/(4\pi)$ in their respective power spectra.}, the vacuum becomes polarized with a refractive index greater than unity~\cite{BialynickaBirula:1970vy,Adler:1971wn,Brezin:1971nd,Tsai:1975iz,Latorre:1994cv,Dittrich:1998fy,Gies:1998sh,Dittrich:2000zu,Marklund:2006my}. As a result, for ultra-relativistic charged particles travelling in such media, the emission of Cherenkov radiation becomes a real possibility. Taking into account the effects of non-linear QED, in the presence of a constant and spatially uniform magnetic field, the Cherenkov power spectrum is given by~\cite{Macleod:2018zcb}
\begin{eqnarray}
&& \frac{dP_{\tiny{\mbox{Cher}}}}{d\omega}=\int^{2\pi}_{0}d\phi \label{eq:chera}
\left(\frac{d^{2}P_{+}}{d\omega d\phi}+\frac{d^{2}P_{-}}{d\omega d\phi}\right), \\
&&\frac{d^{2}P_{\pm}}{d\omega d\phi}=\left(\frac{1}{4\pi}\right)\frac{e^{2}\omega_{\pm}\beta}{2\pi }|\boldsymbol{\epsilon}_{0}\cdot\boldsymbol{\epsilon_{\pm}}|^{2}\sin^{2}\theta_{\pm},\label{eq:cher2}\\
&& \cos\theta_{\pm}\equiv\frac{1}{n_{\pm}\beta}\label{eq:cher}
\end{eqnarray}
where $\beta$ is the velocity of the charged particle which we will take to be electron, $n_{\pm}$ the refractive index and  $\omega_{\pm}$ the modified photon frequency~\cite{BialynickaBirula:1970vy,Adler:1971wn}
\begin{eqnarray}
&& n_{\pm}=(1-\lambda_{\pm}|\mathbf{Q}|^{2})^{-1},\quad \omega_{\pm}=|\k|(1-\lambda_{\pm}|\mathbf{Q}|^{2}),\\
&& \mathbf{Q}\equiv \hat{\k}\times(\hat{\k}\times\mathbf{B})
\end{eqnarray}
with $\mathbf{k}$ being the photon momentum. The coefficients $\lambda_{\pm}$ are derived from the Euler-Heisenberg Lagrangian in the weak field limit~\cite{Schwinger:1951nm}
\begin{equation}
\frac{1}{4}\lambda_{+}=\frac{1}{7}\lambda_{-}=\frac{e^{4}}{360\pi^{2} m^{4}_{e}}.
\end{equation}
In eq.~(\ref{eq:cher}), $\boldsymbol{\epsilon}_{0}$ and $\boldsymbol{\epsilon}_{\pm}$ are the polarization vectors for the Cherenkov radiation and linearly polarized photons in the medium respectively. Their expressions can be found in~\cite{BialynickaBirula:1970vy} but they are not important to us here. Since the refractive index is constant, the spectrum is directly proportional to the photon frequency.

For electrons travelling parallel or anti-parallel to the photon trajectory, the condition for Cherenkov radiation is $\beta>1/n_{\pm}$. For ultra-relativistic electrons, this condition is equivalent to 
$\gamma_{e}^{2}> 1/(2\lambda_{\pm}|\mathbf{Q}|^{2})$. Therefore, in the presence of a strong magnetic field and with the electrons satisfying the stated condition, they will emit both Cherenkov and synchrotron radiation; for convenience, we write the associated $\gamma_{e}$ factor as
\begin{equation}
\gamma^{2}_{e}=X/(2\lambda|\mathbf{B}|^{2}), \quad X>1. \label{eq:gc}
\end{equation}
The power spectrum of the synchrotron radiation is given by~\cite{Schwinger:1949ym}
\begin{equation}
\frac{dP_{\tiny{\mbox{sync}}}}{d\omega}=\left(\frac{1}{4\pi}\right)\frac{3^{3/2}}{4\pi}\frac{e^{2}\gamma_{e}^{4}}{R}\frac{\omega_{0}\,\omega}{\omega^{2}_{c}}\int^{\infty}_{\omega/\omega_{c}}d\eta\, K_{5/3}(\eta)\label{eq:sync}
\end{equation}
where $K_{5/3}(\eta)$ is the modified Bessel function, $R$ the radius of curvature and 
\begin{equation}
\omega_{0}=\frac{e}{E_{e}}|\mathbf{B}|,\quad \omega_{c}=\frac{3}{2}\omega_{0}\gamma_{e}^{3}
\end{equation}
where $E_{e}$ is the energy of the electron. For ultra-relativistic electrons, the radius of curvature can be approximated as $R\approx 1/\omega_{0}$ so that
\begin{equation}
\frac{dP_{\tiny{\mbox{sync}}}}{d\omega}
\approx\left(\frac{1}{4\pi}\right)\frac{1}{\sqrt{3}\pi}\frac{e^{2}\omega_{c}}{\gamma^{2}_{e}}\left[\frac{\omega}{\omega_{c}}
\int^{\infty}_{\omega/\omega_{c}}d\eta\, K_{5/3}(\eta)\right].
\end{equation}
Substituting eq.~(\ref{eq:gc}) into $\omega_{c}$ and taking $\lambda_{\pm}\approx\lambda\sim10^{-32}\,\mbox{Gauss}^{-2}$, we obtain
\begin{equation}
\omega_{c}=8.67X\times10^{23}\,\mbox{eV}\left(\frac{\mbox{Gauss}}{|\mathbf{B}|}\right)\label{eq:wc}
\end{equation}
Above the critical frequency $\omega_{c}$, the synchrotron power spectrum exhibits approximate exponential decay. Therefore, if one extrapolates the photon frequency beyond $\omega_{c}$, the Cherenkov radiation given in eqs.~(\ref{eq:chera}-\ref{eq:cher}) will eventually dominate over its synchrotron counterpart. However, as we will show below, such an extrapolation is problematic because the photon dispersion and refractive index are derived from an effective field theory. Specifically, the extrapolation is constrained by the following inequality~\cite{Narozhny,Ritus} 
\begin{equation}
\chi_{\gamma}=\frac{e}{m^{3}_{e}}[(F\cdot k)^{\mu}(F\cdot k_{\mu})]^{1/2}\lesssim1
\end{equation}
where $F$ is the electromagnetic field strength tensor and $k^{\mu}$ is the photon momentum.
In terms of the critical magnetic field $|\mathbf{B}_{c}|$ and the presence of a spatially constant and uniform magnetic field, we have $\chi_{\gamma}\sim (|\mathbf{B}|/|\mathbf{B}_{c}|) (\omega/m_{e})$ thus yielding an upper-bound to the photon frequency
\begin{equation}
\omega\lesssim \omega_{\tiny{\mbox{max}}}=\frac{|\mathbf{B}_{c}|}{|\mathbf{B}|}m_{e}=2.26\times 10^{19}\,\mbox{eV}
\left(\frac{\mbox{Gauss}}{|\mathbf{B}|}\right).\label{eq:ub}
\end{equation}
Going beyond the bound, the effective field theory becomes unreliable and quantum effects become important. For instance, when $|\mathbf{B}|\sim|\mathbf{B}_{c}|$ and $\omega\gtrsim2m_{e}$, photo-pair production $\gamma\rightarrow e^{+}e^{-}$ dominates~\citep{Adler:1971wn}.

Comparing eqs.~(\ref{eq:wc}) to (\ref{eq:ub}), we find that $\omega_{\tiny{\mbox{max}}}\ll\omega_{c}$. Therefore, we cannot extend the photon frequency associated with Cherenkov radiation beyond $\omega_{c}$ thus verifying our earlier assertion. Therefore,
we may use the following approximation for the synchrotron spectrum~\cite{Rybicki:2004hfl} 
\begin{equation}
\left(\frac{dP_{\tiny{\mbox{sync}}}}{d\omega}\right)_{\omega\ll\omega_{c}}
\approx\left(\frac{1}{4\pi}\right)\frac{4}{3}\left[\frac{e^{2}\omega_{c}}{2^{1/3}\Gamma(1/3)\gamma_{e}^{2}}\right]\left(\frac{\omega}{\omega_{c}}\right)^{1/3}.
\end{equation}
To approximate the maximum Cherenkov power spectrum, we take $|\boldsymbol{\epsilon}_{0}\cdot\boldsymbol{\epsilon}_{\pm}|=1$, set $\beta=1$ and take $\omega_{\pm}\approx\omega$, $\lambda\sim 10^{-32}\,\mbox{Gauss}^{-2}$. These approximations are possible because we are working under the condition $\lambda|\mathbf{B}|^{2}\ll1$. To the leading order in $\lambda|\mathbf{B}|^{2}$, with electrons and photons travelling normal to the magnetic field and $\gamma_{e}^{2}>1/(2\lambda|\mathbf{B}|^{2})$, we get
\begin{equation}
\frac{dP^{\tiny{\mbox{(max)}}}_{\tiny{\mbox{Cher}}}}{d\omega}\approx\left(\frac{1}{4\pi}\right) 4e^{2}\omega\lambda|\mathbf{B}|^{2}.\label{eq:maxC}
\end{equation}
At $\omega=\omega_{\tiny{\mbox{max}}}$ and with $\gamma^{2}_{e}=X/(2\lambda|\mathbf{B}|^{2})$, after some manipulations, we find
\begin{equation}
\left(\frac{dP^{\tiny{\mbox{(max)}}}_{\tiny{\mbox{Cher}}}}{d\omega}\right)\Big{/}
\left(\frac{dP_{\tiny{\mbox{sync}}}}{d\omega}\right)_{\omega\ll\omega_{c}}\approx 2.99\times10^{-3}X^{1/3}.
\end{equation}
Therefore, the Cherenkov radiation becomes dominant when $\gamma_{e}\gtrsim 4.32\times10^{19}\,(\mbox{Gauss}/|\mathbf{B}|)$. 

Even though the Cherenkov radiation dominates for certain ranges of $\gamma_{e}$, actual detections of the Cherenkov radiation from astrophysical sources remains difficult. A common feature of Cherenkov and synchrotron radiation is that both emissions are concentrated in a narrow cone about the path of the electrons. Let $\theta_{\pm}\approx\theta_{\tiny{\mbox{Cher}}}$ and $\theta_{\tiny{\mbox{sync}}}$ be the opening angles for the Cherenkov and synchrotron radiation respectively. For ultra-relativistic electrons, they are given by
\begin{equation}
\theta^{2}_{\tiny{\mbox{Cher}}}\approx 2\lambda|\mathbf{B}|^{2},\quad \theta^{2}_{\tiny{\mbox{sync}}}\approx\frac{1}{\gamma_{e}^{2}}.
\end{equation}
Defining their spectral flux densities to be
\begin{eqnarray}
&&\mathcal{F}_{\tiny{\mbox{Cher}}}(\omega)\equiv\frac{1}{\theta^{2}_{\tiny{\mbox{Cher}}}D^{2}}\frac{dP^{\tiny{\mbox{(max)}}}_{\tiny{\mbox{Cher}}}}{d\omega},\\
&&\mathcal{F}_{\tiny{\mbox{sync}}}(\omega)\equiv\frac{1}{\theta^{2}_{\tiny{\mbox{sync}}}D^{2}}\frac{dP_{\tiny{\mbox{sync}}}}{d\omega}
\end{eqnarray}
where $D$ is the distance between the source and the observer.
At $\omega=\omega_{\tiny{\mbox{max}}}$, the ratio of the flux density is given by
\begin{equation}
\frac{\mathcal{F}_{\tiny{\mbox{Cher}}}}{\mathcal{F}_{\tiny{\mbox{sync}}}}\approx \frac{2.29\times10^{-3}}{X^{2/3}}
\end{equation}
Since $X>1$, the spectral flux density of Cherenkov radiation is at least three orders of magnitude smaller than its synchrotron counterpart.

In summary, the production of Cherenkov radiation in strong magnetic fields is a real and fascinating possibility. However, because $\omega_{\tiny{\mbox{max}}}\ll\omega_{c}$, actual detection of Cherenkov radiation from astrophysical sources remains challenging. It may also be instructive to study photon propagation in strong electric fields or other configurations but it seems to us that the pure magnetic field configuration is the most interesting and physically relevant.

\textit{Acknowledgement -} I would like to thank Chris Gordon and Lance Labun for discussions and reading the manuscript. I would also like to thank the referee for valuable comments. I am grateful to the generous hospitality offered by the Department of Physics and Astronomy at the University of Canterbury where part of this work was completed.



\bibliographystyle{elsarticle-harv} 
\bibliography{Bibliography}






\end{document}